\documentclass[journal]{IEEEtran}
\ifCLASSINFOpdf
   \usepackage[pdftex]{graphicx}
\else
\fi
\usepackage[cmex10]{amsmath}

\begin{document}
%
\title{The Hidden K-edge Signal in K-edge Imaging}
%
%
%

\renewcommand\footnotemark{} 

\author{Christopher~J.~Bateman$^*$,
        Kishore~Rajendran,
        Niels~J.A.~de~Ruiter,
        Anthony~P.~Butler,
        Philip~H.~Butler,
        and~Peter~F.~Renaud}
\thanks{\textit{Asterisk indicates corresponding author}.}
\thanks{$^*$C. J. Bateman  is with the Department of Radiology, University of Otago, Christchurch 8140, New Zealand. (email: christopher.bateman@otago.ac.nz)}
\thanks{K. Rajendran  is with the Department of Radiology, University of Otago, Christchurch 8140, New Zealand. (email: kishore.rajendran@canterbury.ac.nz)}
\thanks{N. J.A. de Ruiter  is with the Department of Radiology, University of Otago, Christchurch 8140, New Zealand. (email: niels.deruiter@canterbury.ac.nz )}
\thanks{A. P. Butler is with the Department of Radiology, University of Otago, Christchurch 8011, New Zealand, and also with the Department of Electrical and Computer Engineering, University of Canterbury, Christchurch 8140, New Zealand, and also with the European Organization for Nuclear Research (CERN), 23 Geneva, Switzerland, and also with MARS Bioimaging Ltd., Christchurch 8140, New Zealand (e-mail: anthony@butler.co.nz).}
\thanks{P. H. Butler is with the Department of Physics and Astronomy, University of Canterbury, Christchurch 8140, New Zealand, and also with the European Organization for Nuclear Research (CERN), 23 Geneva, Switzerland, and also with MARS Bioimaging Ltd., Christchurch 8140, New Zealand (e-mail: phil@butler.co.nz).}
\thanks{P. F. Renaud is with the Department of Mathematics and Statistics, University of Canterbury, Christchurch 8140, New Zealand (email: peter.renaud@canterbury.ac.nz)}%
\maketitle

\begin{abstract}
K-edge imaging is commonly used for viewing contrast pharmaceuticals in a variety of multi-energy x-ray imaging techniques, ranging from dual-energy and spectral computed tomography to fluoroscopy. When looking for the K-edge signal of a specific contrast, by taking measurements either side of the K-edge, it is found that the K-edge is not always observable for low concentrations. We have also observed that the ability to see the K-edge is unit dependent - a K-edge that is not observable in computed tomography (CT) reconstructed linear attenuation units can often be made visible by converting to Hounsfield units.   
\newline \newline
This paper presents an investigation of this K-edge hiding phenomenon. We conclude that if a multi-energy x-ray measurement of any K-edge material contains a signal of any other material, then there will be a positive concentration of that K-edge material below which its K-edge cannot be observed without extracting the K-edge signal through means of basis decomposition. Mathematical descriptions of this limiting minimum concentration is provided in three cases - multi-energy radiographic projection images; and reconstructed multi-energy CT images for both linear attenuation and Hounsfield units. Two important properties of this limiting concentration are provided: the minimum concentration for Hounsfield units is always strictly less than the minimum concentration for linear attenuation; and the minimum concentration for radiographic projections will typically be much larger than the minimum concentration for reconstructed linear attenuation. Finally experimental observation of this phenomenon is presented using data collected by the MARS spectral CT scanner equipped with a CdTe Medipix-3RX camera operating in charge-summing mode, where it is also shown that the K-edge can be recovered through basis decomposition.  
\end{abstract}

\begin{IEEEkeywords}
computed tomography, spectral, multi-energy, dual-energy, K-edge imaging, material decomposition, Medipix.
\end{IEEEkeywords}

%
\IEEEpeerreviewmaketitle

\section{Introduction}
%
%
%
%

 

\IEEEPARstart{I}n medical x-ray imaging it is common practice to use contrast pharmaceuticals containing highly attenuating heavy elements such as iodine, barium or gadolinium to highlight parts of the anatomy to enable the diagnosis of many conditions \cite{1977Winkler,1978Drayer,1978Foley,2014Kong,2010Anderson}. Each of these heavy elements have a characteristic jump in their attenuation (opacity) of x-rays with energies above the binding energy of their K-shell electrons. This jump, also known as the K-edge, is used to identify specific heavy elements in multi-energy x-ray imaging techniques such as digital subtraction angiography (DSA) \cite{1981Brody}, dual-energy computed tomography (dual-energy CT) \cite{2013Zhang} and spectral computed tomography (spectral CT) \cite{2008Schlomka,2010Anderson}. The measurement of a specific heavy element's K-edge in multi-energy x-ray imaging is widely known as K-edge imaging. 
\newline \newline
K-edge imaging requires that measurements be taken at x-ray energies (or effective energies) either side of the respective K-edge. This can be achieved in three ways (in order of decreasing K-edge resolution): using two different monochromatic x-ray beams generated by a synchrotron; measuring a polychromatic x-ray spectrum with an energy discriminating photon counting detector (such as those found in the Medipix detector family) \cite{2011Walsh}; or measuring two different polychromatic x-ray beams (with mean energies either side of the K-edge) using integrating detectors \cite{2009Xu,2006Flohr}. The latter of these three methods is the typical acquisition used for dual-energy CT.    
\newline \newline
The general K-edge imaging problem can be stated as ``given that a heavy element is present in an object and that a multi-energy acquisition has been obtained for its respective K-edge, which parts of the object (or anatomy) contain the heavy element? and which parts do not?''. Image processing techniques commonly used to solve this problem, which are sensitive to the K-edge signal, include basis decomposition (also known as material decomposition) \cite{1976Alvarez,2011Le,2004Firsching} and image subtraction \cite{1977Kruger,2012Pani,2007Sarnelli,1977Riederer}. Basis decomposition splits a signal up into contributions from its various constituents, which enables the component of the signal representing the heavy element (and its K-edge) to be analysed separate to the remainder of the signal. K-edge subtraction imaging subtracts images representing acquisitions at x-ray energies either side of the respective K-edge. The result of this subtraction is that any regions containing significant levels of that heavy element will have a different sign (-/+) to other materials present in the image (such as tissue and bone), making identification of the heavy element trivial. It should also be noted that looking by eye for the K-edge jump in an attenuation versus photon energy plot (for example what can be obtained from the NIST XCOM database \cite{NIST}) is equivalent to K-edge subtraction. The main difference between these two types of techniques is that subtraction imaging does not decouple the K-edge material signal from other materials that are typically present. This provokes an interesting question - how sensitive are K-edge subtraction imaging techniques for identifying heavy elements that are in composition with other materials (i.e. tissue + contrast, multiple contrasts, or even other components of the given pharmaceutical).      
\newline \newline
The work presented in this paper investigates the theoretical minimum concentration at which heavy elements can be identified using K-edge subtraction imaging. It was found that this minimum concentration depends on the x-ray energy ranges used in the measurement, the K-edge of the heavy element, and all other materials coupled to the signal of the heavy element. Formulae for the minimum concentrations for multi-energy x-ray data are presented for three cases: x-ray projection data, CT reconstructed Hounsfield units, and CT reconstructed linear attenuation units. Of these three cases, we show that reconstructed Hounsfield units have the best K-edge detectability for subtraction techniques whereas projection data has the worst K-edge detectability.     
\newline \newline
Experimental observation of this minimum concentration is provided using a MARS multi-energy CT scanner equipped with a CdTe Medipix-3RX energy-discriminating photon-counting detector operating in charge summing mode. Clinically available iodine and gadolinium pharmaceuticals (Visipaque\texttrademark $\,$and Omniscan\texttrademark ) were scanned at various levels of dilution. Obscuration of the K-edge in narrow-bin reconstructed effective linear attenuation units was observed at concentrations $\sim 20\,$mg/ml with the selected scan parameters. Basis decomposition was also performed on the dilutions found to be below the minimum concentration for K-edge subtraction. This was used to confirm that the K-edge was indeed present in the decoupled heavy element signal for those concentrations.  
\newline \newline
It should be noted that this minimum concentration is a limitation that is independent of the level of measurement noise (with the exception of noise in the energy resolution of the measurement system). This means that it is conceivable that this minimum detectable concentration can be considerably higher than what would be expected from typical measurement noise and systematic uncertainties.   

\section{background}

\subsection{The K-edge in Multi-energy X-ray Imaging}
The K-edge is a feature found in the x-ray attenuation spectra of most elements. It is described by a jump in attenuation for x-rays which have an energy greater than the binding energy of the K-shell electrons. Contrast pharmaceuticals typically contain elements with K-edge energies in the human imaging range ($30-140\,$keV). The effect of these K-edge features can be explicitly visible in multi-energy X-ray measurements. 

\begin{figure}[!h]
\begin{center}
\includegraphics[width=0.45\textwidth]{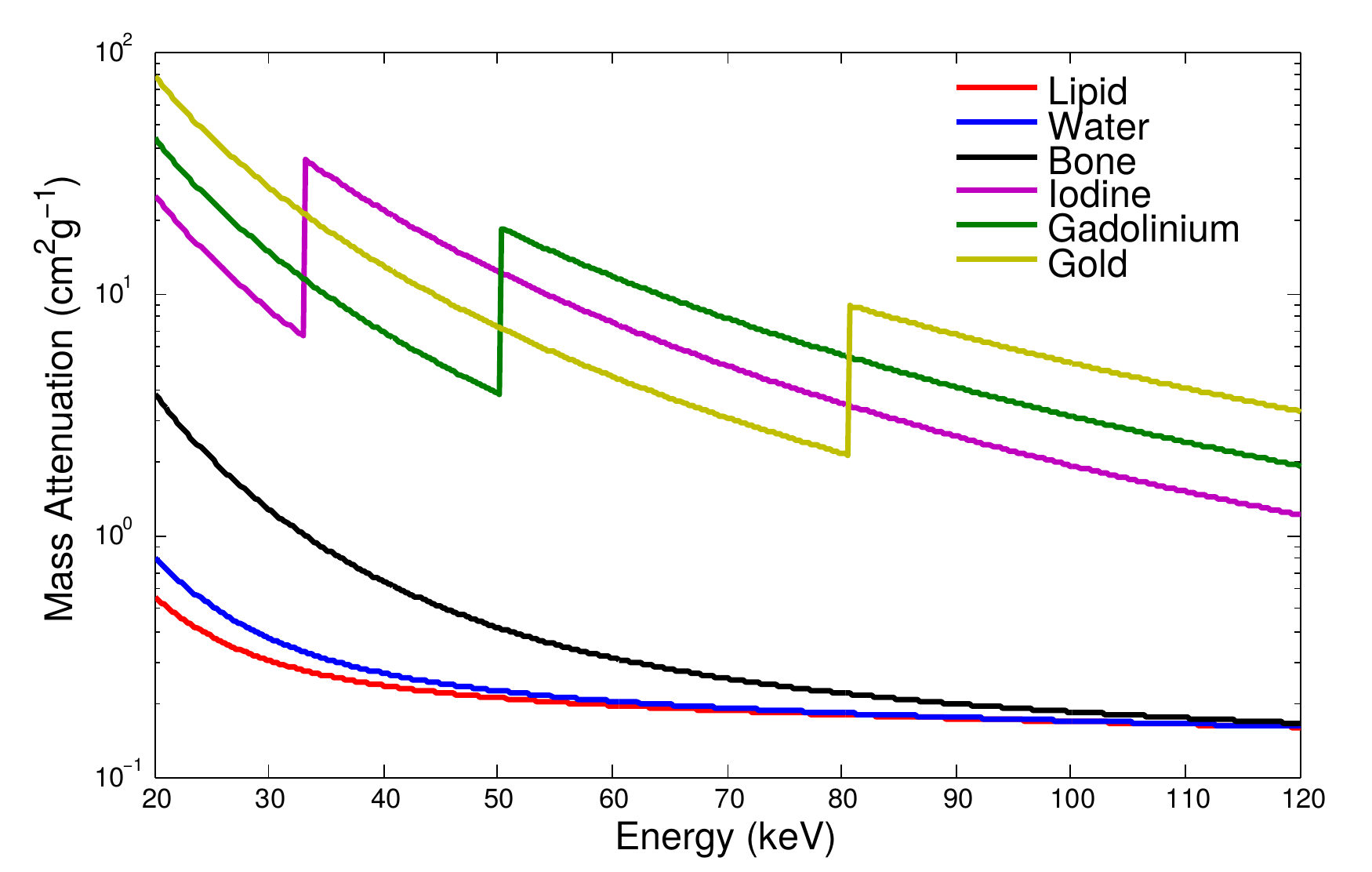}
\end{center}
\caption{Mass attenuation of various materials in the diagnostic imaging range showing K-edges for heavy elements Iodine, Gadolinium and Gold.}
\label{Fig:AttPlot}
\end{figure}  

\subsection{K-edge Subtraction Imaging}
K-edge subtraction imaging is the digital subtraction of two x-ray images acquired from effective energy ranges either side of a K-edge. Because the attenuation of all other present materials (bone, tissue, etc...) decrease with increasing x-ray energy, the subtraction effectively removes other materials while enhancing the attenuation jump due to the K-edge. This results in an image showing only where the K-edge pharmaceutical is distributed.

\subsection{Basis Decomposition}
Basis decomposition is a technique from linear algebra which is most simply stated as the process of solving systems of linear equations. It was first used for material analysis in multi-energy x-ray imaging by Alvarez and Macovski in 1976 \cite{1976Alvarez}. They used basis decomposition to split dual-energy CT x-ray projections into their photoelectric and Compton scattering contributions. Many variants of this method have since been formulated for identifying contributions of individual materials (also known as material decomposition).

A common application of basis decomposition is to identify the contrast pharmaceutical component of a scan, then subtract that component to achieve an equivalent contrast-free image without performing an extra scan. Coincidently this is also sometimes referred to as K-edge subtraction, however it is not the subtraction technique focused on in this paper.   
\section{The Hidden K-edge Effect}
\begin{figure}[!h]
\begin{center}
\includegraphics[width=0.45\textwidth]{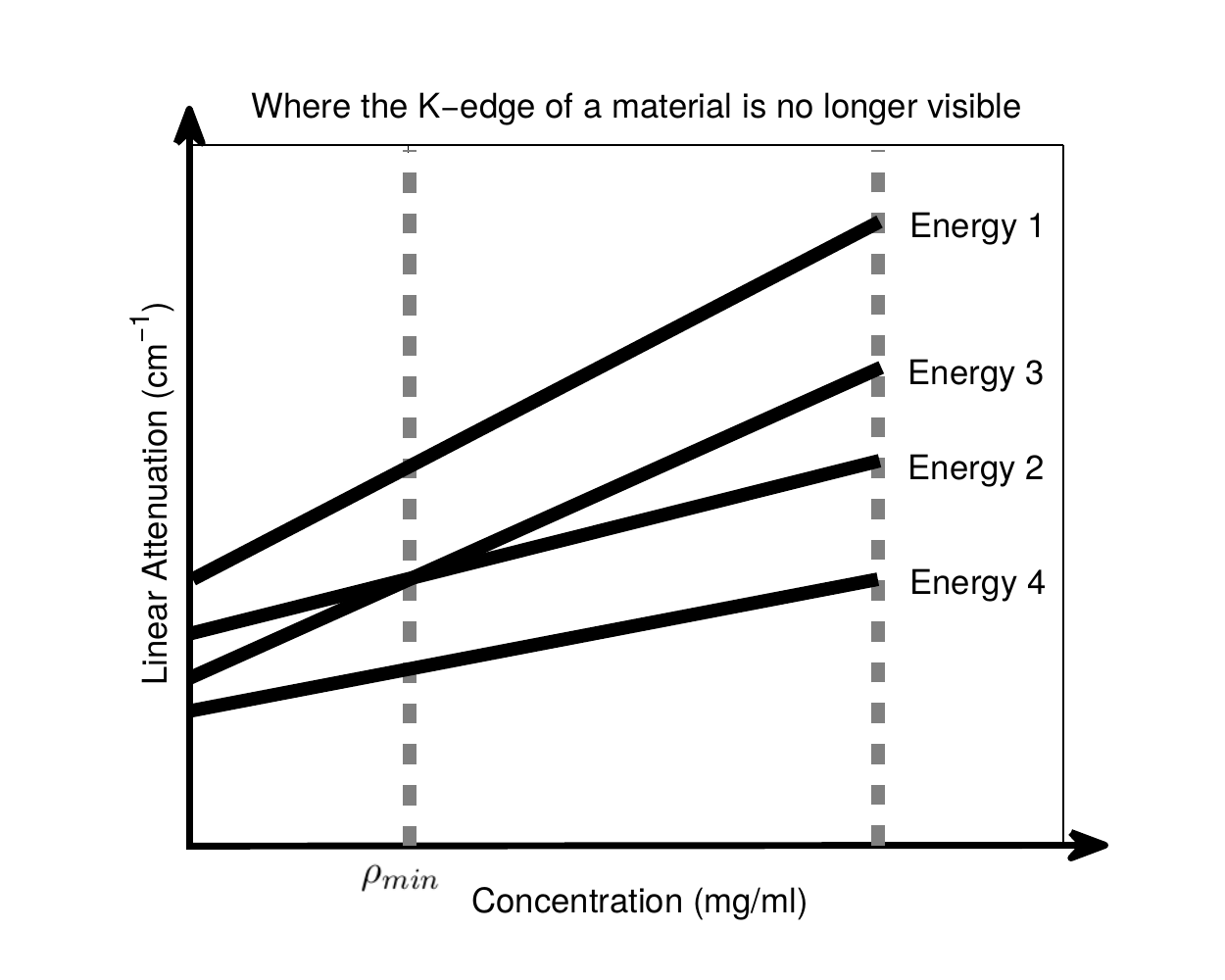}
\end{center}
\caption{The attenuation of a material in solution is linearly proportional to concentration of the solution. Taking energy 2 and energy 3 to be measurements of the K-edge, the concentration $\rho_{min}$ is where the two energies have the same attenuation. For concentrations above $\rho_{min}$ the K-edge is visible and below the K-edge is not visible.}
\label{Fig:K-edge_Att}
\end{figure}

The hidden K-edge (or K-edge hiding) effect is a phenomenon in K-edge imaging where the K-edge of the respective high Z material cannot be observed at low concentrations independent of the degree of noise present in the data. K-edge hiding arises from the linearity properties of x-ray attenuation for composite materials
\begin{equation}
\mu(E) = \sum_m^{materials} \rho_m \left( \frac{\mu}{\rho} \right)_m(E)
\label{Eqn:linearity}
\end{equation}
where $\mu(E)$ is the linear attenuation of the composite material for x-rays of energy $E$, and $\rho_m$ and $(\mu/\rho)_m$ are the density/concentration and mass attenuation respectively for the $m^{th}$ material of the composition. Also note that for the remainder of the paper the energy dependence of the attenuation terms will be written as a superscript, $\mu(E) \rightarrow \mu^E$, to reduce verbose use of parentheses.     
\newline \newline
As a thought experiment, let the K-edge material be in solution (with water). If the K-edge material is at high enough concentration we will see a K-edge in a multi-energy x-ray measurement. If we reduce the concentration to zero then there is only water left and there is no K-edge in the measurement. Due to the continuity properties of Eqn. (\ref{Eqn:linearity}) there must be a concentration $\rho_{min}$ below which the K-edge is not visible in the measurement - see Fig. \ref{Fig:K-edge_Att}. This concentration is found where the solution has the same attenuation for the two energies measuring the K-edge. It should be noted that for this example we can replace water with any surrogate material or combination of materials (e.g. bone, tissue, blood, another K-edge pharmaceutical, or even the non K-edge constituents of these pharmaceuticals). 
\newline \newline 
This remainder of this section presents formulae derived for calculating the minimum concentration in three different cases: reconstructed multi-energy CT images in both linear attenuation $\rho_{min}^{LA}$ and Hounsfield units $\rho_{min}^{HU}$; and for multi-energy radiographic projections which needs to consider minimum material mass over x-ray path-length $M_{min}^{Proj}$ rather than concentration (as concentrations can vary considerbly over long projection distances). All derivations consider monochromatic x-ray measurements.

\subsection{The Hidden K-edge in Reconstructed Linear Attenuation.}
The linear attenuation of x-rays of energy $E$ through a voxel containing a K-edge material $X$ and another material $m$ with concentrations $\rho_X$ and $\rho_m$ respectively is given by
\begin{equation}
\mu^E = \mu_m^E + \rho_X \left(\frac{\mu}{\rho}\right)_X^E
\label{Eqn:linatt}
\end{equation} 
where $\mu_m^E = \rho_m (\mu / \rho)_m^E \approx constant$ for low concentrations of $X$. 
\newline \newline
Taking Eqn \ref{Eqn:linatt} for two different x-ray energies $a$ and $b$ (either side of the K-edge) we can find a $\rho_X$ such that both energies have the same linear attenuation. This is equivalent to finding the intersection of the lines for energies 2 and 3 in Fig. \ref{Fig:K-edge_Att}. This concentration is given by
\begin{equation}
\rho_{min}^{LA} = \frac{\mu_m^a - \mu_m^b}{\left(\frac{\mu}{\rho}\right)_X^b - \left(\frac{\mu}{\rho}\right)_X^a}
\label{eq:LA}
\end{equation} 

\subsection{The Hidden K-edge in Reconstructed Hounsfield Units.}
Despite linear attenuation conforming to the SI unit convention ($m^{-1}$ or more commonly $cm^{-1}$), it is standard practice in the field of CT to use Hounsfield units (HU) which defines the value of water to $0$ and air to $-1000$. 
\begin{equation}
HU = 1000 \times \left( \frac{\mu - \mu_{water}}{\mu_{water} - \mu_{air}} \right)
\end{equation}
It therefore deserves special mention in regards to the K-edge hiding effect. Eqn. \ref{Eqn:linatt} can be redefined in terms of HU (for a specific energy $E$) as
\begin{equation}
\frac{HU^E}{1000} = \frac{\mu_m^E}{\mu_{water}^E} + \rho_X \frac{\left(\frac{\mu}{\rho}\right)_X^E}{\mu_{water}^E}
\end{equation} 
\newline \newline
Using the same logic that was used for linear attenuation, we can show that two energies $a$ and $b$ on either side of the K-edge have the same HU when the K-edge material has the concentration
\begin{equation}
\rho_{min}^{HU} = \frac{\mu_m^a \mu_{water}^b - \mu_m^b \mu_{water}^a}{\left(\frac{\mu}{\rho}\right)^b_X \mu_{water}^a - \left(\frac{\mu}{\rho}\right)^a_X \mu_{water}^b}
\label{eq:HUth}
\end{equation}
It should be noted that even though $\rho_{min}^{HU}$ is zero by definition when the K-edge material is in water (and only water!), this is not achievable in any practical application. K-edge pharmaceuticals, when in the body, are present with blood and other tissues; and the high Z atoms contributing the K-edge are typically not the only constituent of the respective pharmaceutical.      

\subsection{The Hidden K-edge in Radiographic Projections.}
To analyse the hidden K-edge effect in radiographic projections we are required to explicitly use the Beer-Lambert law in the form
\begin{equation}
I^E = I_o^E \times \exp(- \mu^E \ell)
\end{equation}
where $I_o^E$ and $I^E$ is the x-ray beam intensity at energy $E$ before and after the beam passes through the object, $\mu^E$ from Eqn. \ref{Eqn:linatt} and $\ell$ is the total path-length through the object. 
\newline \newline
Since the x-ray beam must pass a considerable distance through the object it does not make sense to talk in terms of concentrations for radiographic projections. Instead we will use total mass of material that the beam passes through. If the beam passes through a mass $M_m$ of other materials then by equating the transmitted beam fraction (or normalized projection) $I/I_o$ for each energy $a$ and $b$, the minimum mass $M_{min}^{Proj}$ of K-edge material $X$ the beam needs to pass through to hide the K-edge is given by
\begin{equation}
M_{min}^{Proj} = M_m \frac{ \left(\frac{\mu}{\rho}\right)_m^a - \left(\frac{\mu}{\rho}\right)_m^b }{\left(\frac{\mu}{\rho}\right)_X^b - \left(\frac{\mu}{\rho}\right)_X^a}
\label{Eqn:proj}
\end{equation}  
\newline \newline
If we take limit the size of the object to that of a CT reconstructed voxel then Eqn. \ref{Eqn:proj} (for projections) reduces to Eqn. \ref{eq:LA} (for reconstructed linear attenuation) as expected.
\newline \newline
It should also be noted that if this derivation is done with the equivalence $I^a = I^b$ (i.e. raw projections without normalization), then the numerator of Eqn. (\ref{Eqn:proj}) has the extra term $-(V/\ell)\times \ln(I^a_o/I^b_o)$ added to it. In this case $V$ is the volume of the object the x-ray beam passes through.  

\section{Properties of the hidden K-edge.}
\subsection{Comparing Hounsfield Units to Linear Attenuation}
Here we show one of the consequences of converting reconstructed multi-energy CT data from linear attenuation to Hounsfield units: K-edges can always be observed at lower concentrations in Hounsfield units compared to linear attenuation. To show this relationship we start by equating the minimum concentration for each unit.
\begin{multline}
\rho^{HU}_{min} = \rho^{LA}_{min} \times \left[\frac{\left(\frac{\mu}{\rho}\right)^b_{X} - \left(\frac{\mu}{\rho}\right)^a_{X}}{\mu^a_{m} - \mu^b_{m}}\right] \times \hdots \\ \hdots \times \left[ \frac{\mu^{a}_{m} \mu^{b}_{water} - \mu^{b}_{m} \mu^{a}_{water}}{\left(\frac{\mu}{\rho}\right)^{b}_{X}\mu^{a}_{water} - \left(\frac{\mu}{\rho}\right)^{a}_{X}\mu^{b}_{water}}\right]
\label{Eqn:LA-HU}
\end{multline}
Without loss of generality energy $a$ is assumed to be less than energy $b$. For water $\mu^b < \mu^a$, and if the K-edge is being imaged then $(\mu/\rho)_X^b > (\mu/\rho)_X^a$. Using these properties, Eqn. \ref{Eqn:LA-HU} can be reduced to the following inequality.
\begin{equation}
\rho^{HU}_{min} < \rho^{LA}_{min}
\end{equation}

\subsection{Comparing Linear Attenuation to Radiographic Projections}

Comparing the K-edge observation limits for linear attenuation and radiographic projections is trivial (because the formulae for each are virtually the same). Dividing Eqn. \ref{eq:LA} by Eqn. \ref{Eqn:proj} gives the relationship 
\begin{equation}
\frac{M^{proj}_{min}}{M_m} = \frac{\rho^{LA}_{min}}{\rho_m}
\end{equation}
This relationship says that, at the K-edge observation boundary, the ratio of K-edge material mass to non-K-edge material mass in a reconstructed voxel (linear attenuation) is the same as for a radiographic projection along the path-length the respective x-ray beam travels. 
\newline \newline 
The x-ray beam will typically pass through much more tissue in a radiographic projection than in a single reconstructed voxel (in most applications). The consequence of this is that higher concentrations of contrast pharmaceuticals are required to observe K-edges in multi-energy radiographic projection images than in multi-energy CT reconstruction images.

\subsection{Positive, Negative, and Zero Minimum Concentrations}
\begin{figure}[!h]
\begin{center}
\includegraphics[width=0.5\textwidth]{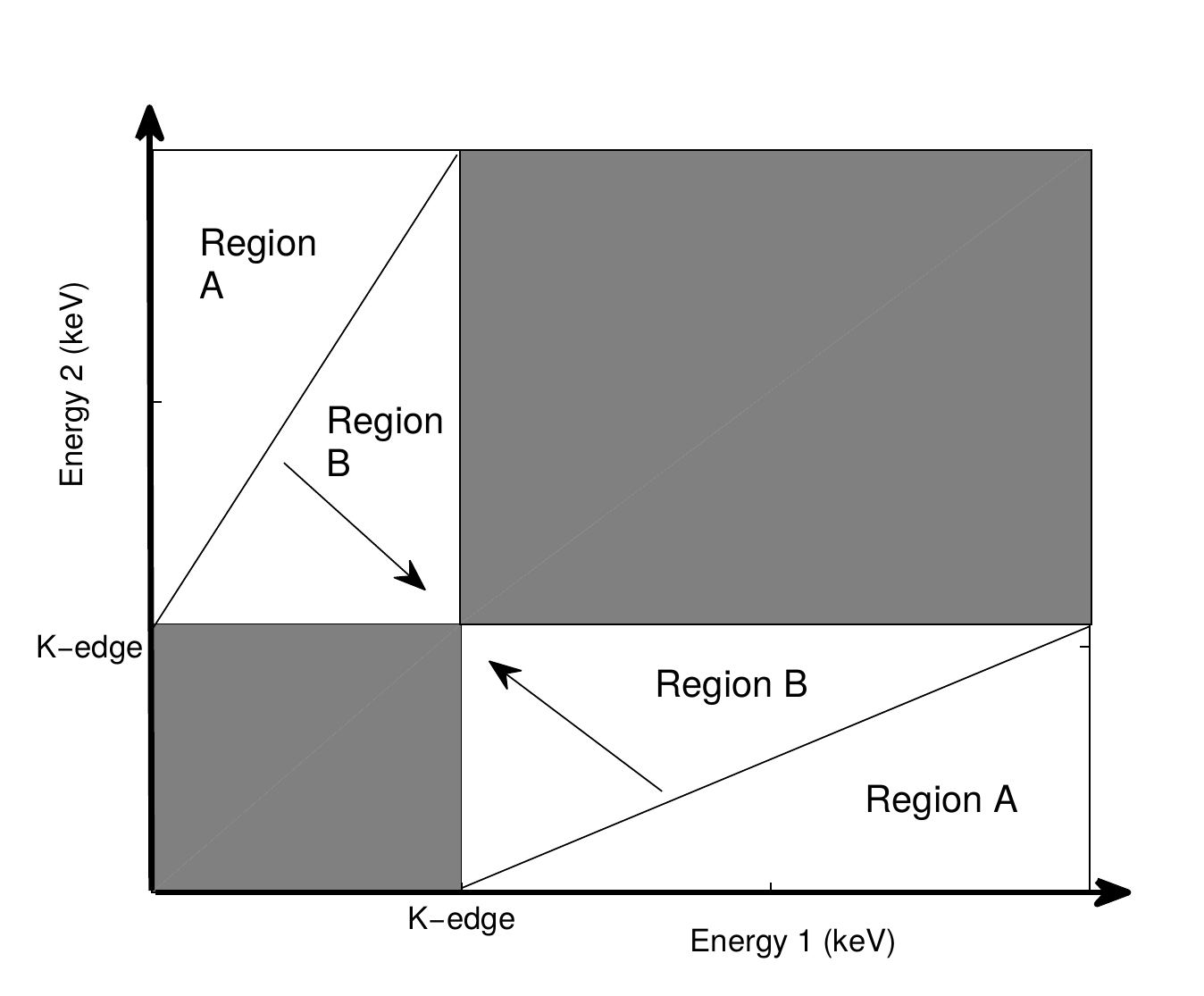}
\end{center}
\caption{Minimum concentration trends (for any combination of two energies) for any K-edge material in solution. The K-edge is only ever visible in Region B (grey region and Region A have negative minimum concentration). Arrows point in direction of decreasing minimum concentration.}
\label{fig:Contour}
\end{figure}

A summary of when the minimum concentration is positive or negative is presented in Fig. \ref{fig:Contour}. The concentration is only ever negative when minimum concentration formula are used for non-K-edge imaging data (which violates the assumption that K-edge imaging is being performed). When K-edge imaging is being performed the minimum concentration will always strictly positive, except in the following two cases:
\begin{itemize}
\item For reconstructed Hounsfield units when the K-edge material and water are the only materials present, due to the definition of Hounsfield units.
\item When there is no other material present than the K-edge material, which violates the primary assumption that the K-edge material is in composition with another materials.
\end{itemize}

\newpage
\section{Experimental Observation of the Hidden K-edge Effect Using MARS}
This section shows that the hidden K-edge effect is observable in reconstructed data collected using a MARS scanner. In addition, it is also shown that the hidden K-edge signal can be recovered by using basis decomposition to extract the mass  attenuation coefficient representation of the individual K-edge material.
\subsection{Methodology}
Various dilutions of common K-edge contrast pharmaceuticals Omniscan\texttrademark $\,$ and Visipaque\texttrademark $\,$ (containing gadolinium and iodine respectively) were scanned in a phantom using a MARS scanner equipped with a Medipix CdTe3RX detector bump bonded at 110$\,\mu$m and operated in charge summing mode. 
\subsubsection{Phantom}
\begin{figure}[!h]
\begin{center}
\includegraphics[scale=0.8]{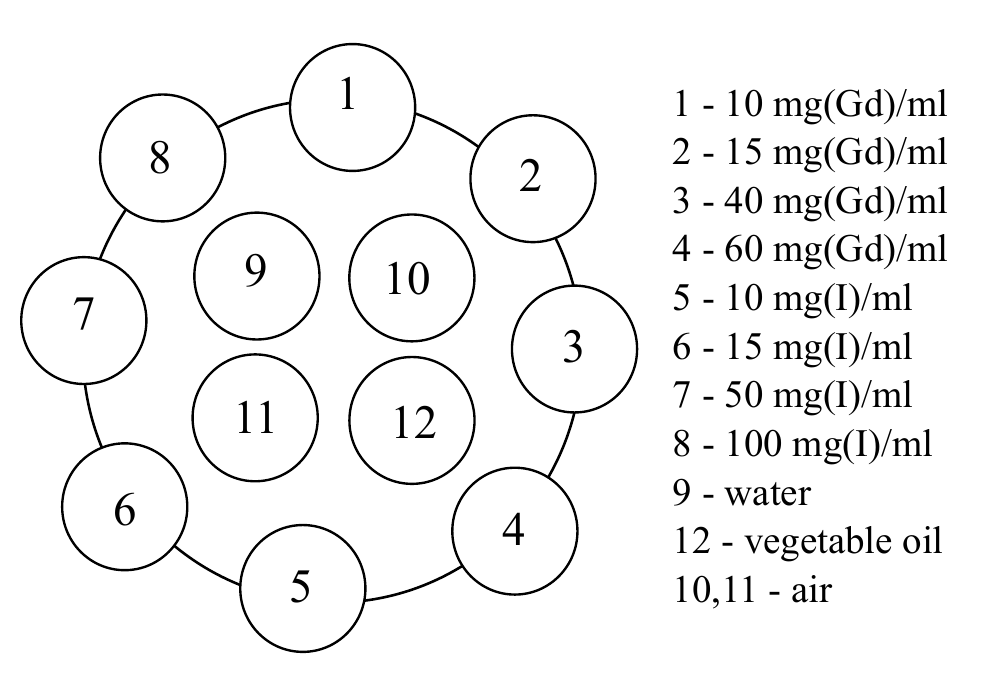}
\end{center}
\caption{Phantom capillary contents. Omniscan\texttrademark $\,$dilutions are used for gadolinium solutions and Visipaque\texttrademark $\,$dilutions are used for iodine solutions. }
\label{fig:phantom}
\end{figure}

The phantom consists of a solid polymethyl methacrylate (PMMA) cylinder with four capillaries drilled in the middle and eight around the outer edge. Samples contained in small PCR tubes were inserted into the capillaries. Solutions used include Visipaque\texttrademark $\,$$100$, $50$, $15$, and $10\,$mg(I)/ml, Omniscan\texttrademark $\,$$60$, $40$, $15$ and $10\,$mg(Gd)/ml, water and vegetable oil (see Fig. \ref{fig:phantom}).   

\begin{figure}[!h]
\begin{center}
\includegraphics[scale=0.9]{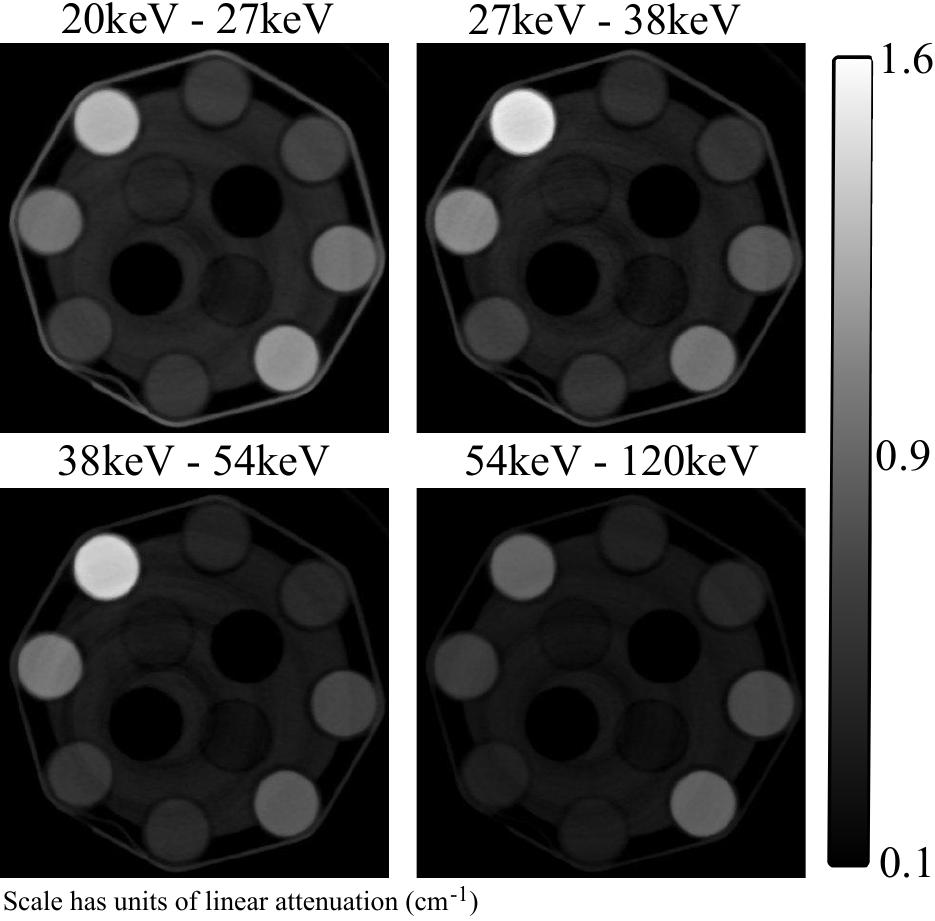}
\end{center}
\caption{Reconstructed images of phantom scan, scale is in units of linear attenuation (cm$^{-1}$). Capillary contents described in Fig. \ref{fig:phantom}. Note that scale has been chosen to be in the units which hide the K-edge at higher concentrations. Fig. \ref{fig:IodExp} and Fig. \ref{fig:GdExp} help show which capillaries have their K-edge hidden.}
\label{fig:Rec}
\end{figure}  

\subsubsection{Scan Parameters and Data Processing}
The scan was acquired using a tungsten x-ray tube operated at $120\,$kVp, $35\,\mu $A, with $1.8\,$mm Al equivalent filtration. A source to object distance of $160\,$mm and an object to detector distance of $100\,$mm was used. Each projection was taken with an exposure time of $100\,$ms. The energy thresholds were chosen to be $20$, $27$, $38$, and $54\,$keV. A total of 980 projections were obtained. Each energy bin was dark-field corrected with 34 dark-field frames and flat-field normalized with 980 open beam frames. Narrow energy bins were reconstructed into effective linear attenuation using an algebraic reconstruction technique (Fig. \ref{fig:Rec}).
\newline \newline
The linear attenuation for each dilution was approximated by taking the median of $53,781$ reconstructed voxels in their respective regions. To show that a hidden K-edge can be recovered by decoupling the contrast attenuation from the water attenuation, the mass attenuation coefficients for each contrast were calculated by solving systems of equations formulated with Eqn. \ref{Eqn:linatt} which used only concentrations which had a hidden K-edge. The systems of equations were solved using the Moore-Penrose pseudo-inverse.     

\subsection{Results}
\begin{figure}[!h]
\begin{center}
\includegraphics[scale=0.65]{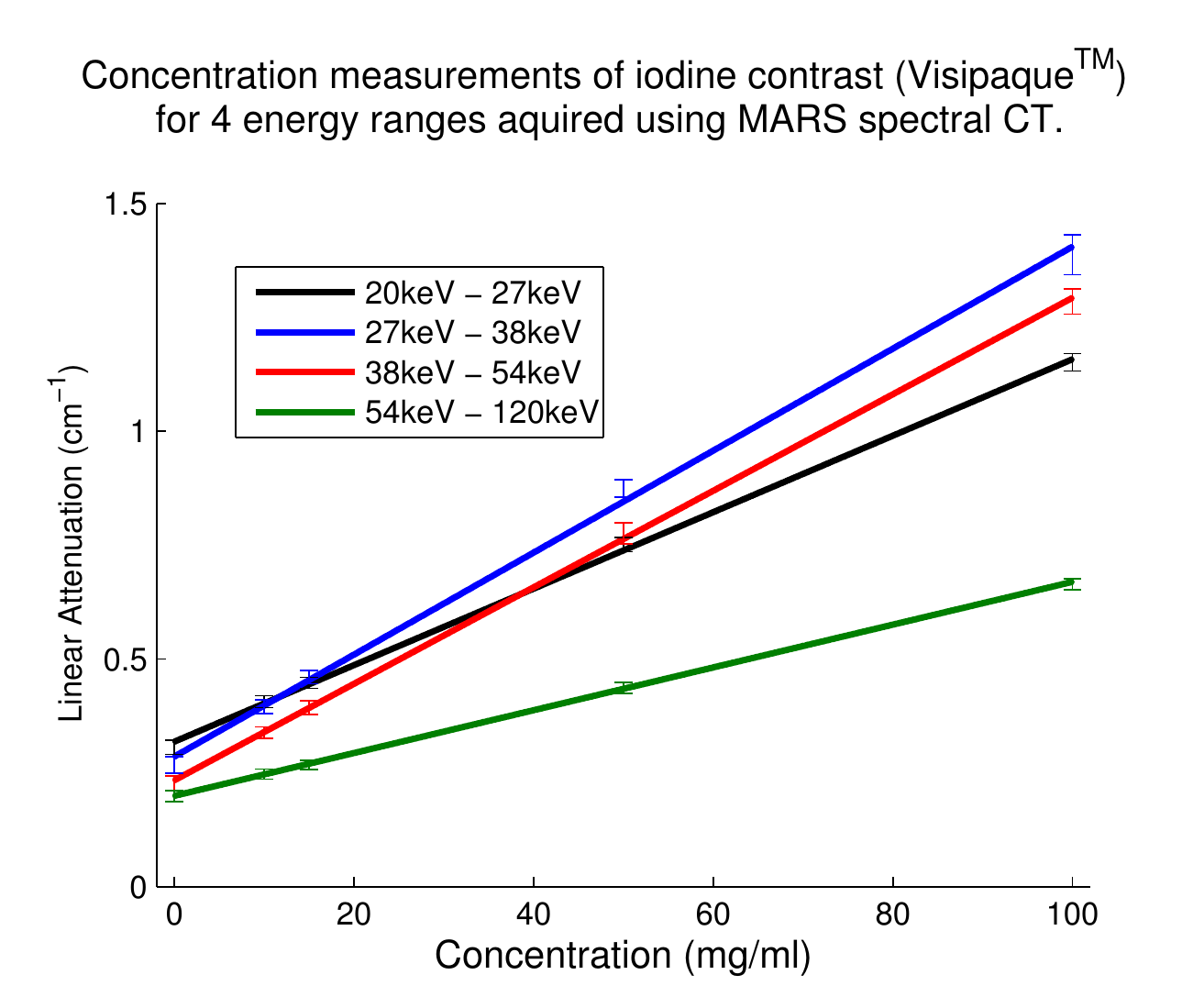}
\caption{Visipaque\texttrademark $\,$has its K-edge span two of the measured energy ranges, each with a different minimum concentration, $38.6\,$mg(I)/ml and $11.7\,$mg(I)/ml. Error bars represent one standard deviation.}
\label{fig:IodExp}
\end{center}
\end{figure}

\begin{figure}[!h]
\begin{center}
\includegraphics[scale=0.65]{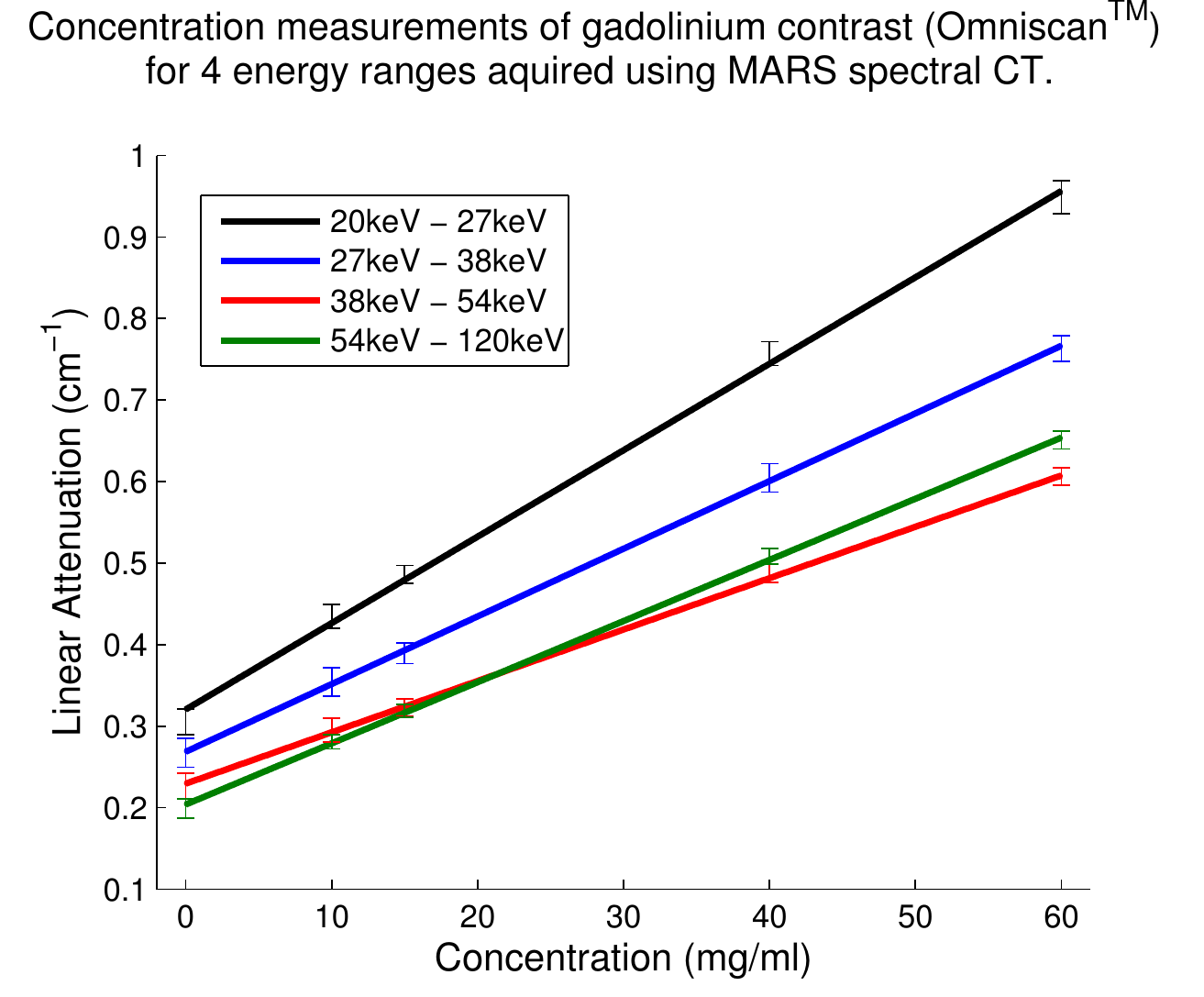}
\caption{Omniscan\texttrademark $\,$was found to have a minimum concentration of $21.3\,$mg(Gd)/ml for the selected energy ranges. Error bars represent one standard deviation.}
\label{fig:GdExp}
\end{center}
\end{figure}

\begin{figure}[!h]
\begin{center}
\includegraphics[scale=0.65]{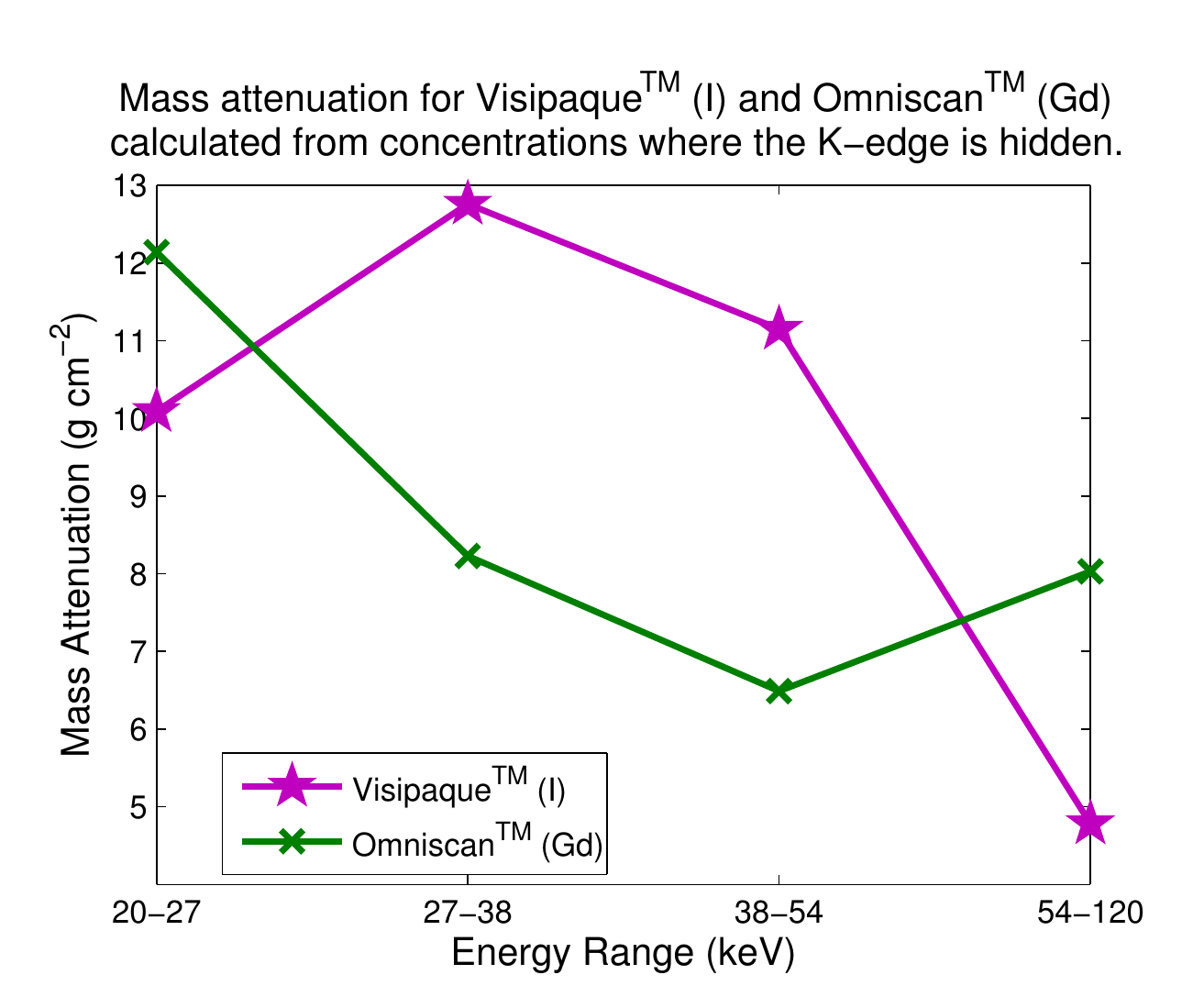}
\caption{Mass attenuation of Visipaque\texttrademark $\,$and Omniscan\texttrademark $\,$calculated from concentrations below the calculated minimum concentration. The removal of water from the signal has made the K-edges visible.}
\label{fig:massAtt}
\end{center}
\end{figure}

The equivalent of Fig. \ref{Fig:K-edge_Att} was produced from the phantom data for both Omniscan\texttrademark $\,$ and Visipaque\texttrademark $\,$(Fig. \ref{fig:IodExp} and \ref{fig:GdExp}). Both contrasts were identified to have positive minimum concentrations for K-edge observation.
\newline \newline
 Visipaque\texttrademark $\,$was found to have the increase in attenuation, due to the K-edge, in the two middle energy bands (which is in agreement with theoretical mass attenuation coefficients for the selected energy ranges). The two minimum concentrations were found: between $20-27\,$keV and $27-38\,$keV the minimum concentration was found to be $11.7\,$mg(I)/ml; and between $20-27\,$keV and $38-54\,$keV the minimum concentration was found to be $38.6\,$mg(I)/ml. Omniscan\texttrademark $\,$was found to have a single minimum concentration between $38-54\,$keV and $54-120$\,keV of $21.3\,$mg(Gd)/ml.   
\newline \newline
The mass attenuation coefficients for Omniscan\texttrademark $\,$and Visipaque\texttrademark $\,$were calculated using concentrations below the observed minimum concentration. It is seen in Fig. \ref{fig:massAtt} that the K-edges were recovered when the contrasts were decoupled from water in this way.   
 
\section{Discussion}
The hidden K-edge effect is the obscuration of a material's K-edge by another material during K-edge imaging. This effect will occur in any multi-energy x-ray measurement of a K-edge material, both in radiographic projections and tomographic reconstructions. The minimum concentration for observing a K-edge using image subtraction techniques depends on: the K-edge material; all other materials the x-ray beam passes through; the x-ray energies chosen to measure the K-edge; and whether the image subtraction is done on projection or reconstruction images. These properties result in the theoretical minimum concentration for K-edge observation being independent of the level of noise present in measured data.  
\newline \newline
An interesting property of the hidden K-edge effect is that K-edge materials can be identified at lower concentrations in reconstructed multi-energy CT images which are scaled to Hounsfield units from linear attenuation. This advantage however is limited only to multi-energy CT imaging modalities. Fluoroscopy type imaging modalities, which display radiographic projections, are the most affected by the hidden K-edge effect typically due to having a greater ratio of tissue to contrast along the x-ray beam path-length than in a reconstructed voxel. 
\newline \newline
The expressions describing the minimum concentrations in this chapter have been derived assuming that monochromatic x-ray beams are used for the K-edge imaging measurements. Despite this simplification, experiments verify that the hidden K-edge effect indeed extends to polychromatic x-ray beams. The experimental results presented in this paper confirms that the hidden K-edge does indeed generalize to polychromatic measurements.   
\newline \newline
Using the experimental results in this paper it is shown that basis decomposition techniques are still able to identify the K-edge for solutions below the minimum concentration. This is because basis decomposition separates the components of the signal belonging to each material, essentially removing the parts of the signal hiding the K-edge. Because of this basis decomposition techniques are more applicable for identifying low concentration K-edge materials than subtraction techniques. The K-edge hiding effect however does not offer any insight into the minimum concentration of K-edge materials detectable by basis decomposition techniques.

\appendices


\section*{Acknowledgment}
This project was funded by Ministry of Business, Innovation and Employment (MBIE), New Zealand under contract number UOCX0805. The authors would like to thank all members of MARS-CT project, the Medipix2 collaboration, and the Medipix3 collaboration. In particular we acknowledge the CERN based designers Michael Campbell, Lukas Tlustos, Xavier Llopart, Rafael Ballabriga and Winnie Wong, and the material scientists Alex Fauler, Simon Procz, Elias Hamann, Martin Pichotka, and Michael Fiederle from Freiburger Materialforschungszentrum (FMF) and X-ray Imaging Europe GmbH. We also thank Graeme Kershaw and Joe Healy, University of Canterbury for preparing components of the phantom.


\ifCLASSOPTIONcaptionsoff
  \newpage
\fi



%


\bibliographystyle{IEEEtran}
\bibliography{references}

%








\end{document}